\documentclass[preprintnumbers,twocolumn,amsmath,amssymb]{revtex4-2}

\begin{document}

\preprint{TU-1235}
\preprint{May, 2024}

\title{CPT, Majorana fermions, and particle physics beyond the Standard Model}
\author{Ken-ichi Hikasa}
\affiliation{Division for Interdisciplinary Advanced Research and Education, 
Advanced Graduate School, Tohoku University, Sendai 980-8578, Japan}
\affiliation{Institute of Liberal Arts and Sciences, Institute for Excellence in Higher Education, Tohoku University, Sendai 980-8576, Japan}
\affiliation{Department of Physics, Tohoku University, Sendai 980-8578, Japan}

\begin{abstract}%
After reviewing charge conjugation 
and the CPT theorem, we define Majorana fermions and clarify 
the relationship of Majorana, Weyl, and Dirac fields.  
Appearance of Majorana fermions in various scenarios of physics 
beyond the Standard Model is discussed, including neutrino masses, 
baryon asymmetry of the universe, grand unified theories, and supersymmetry.  
\end{abstract}

\maketitle

\section{Antiparticles}

Our first encounter with antiparticles was in 1932, when Anderson discovered 
positrons in cosmic rays \cite{Anderson}.  
Their existence had been anticipated by Dirac \cite{Dirac} in interpreting 
negative energy solutions of Dirac equation.  

Any particle has its own antiparticle.  Antiparticle is characterized by 
possessing the same mass and lifetime as the particle, while additive quantum 
numbers such as electric charge are opposite in sign.  
If a particle has no nonzero quantum numbers, it is possible that the particle 
is its own antiparticle.  Photon is the oldest known such particle.  

When muons were discovered \cite{muon} in cosmic rays in 1937, some of them 
possessed plus, the others minus electric charges.  The two charge states 
are particles and antiparticles of each other.  
As for mesons found in 1940's and 50's, pions are in three charge states, 
among which $\pi^+$ and $\pi^-$ form a particle-antiparticle pair, 
and $\pi^0$ is its own antiparticle.  Kaons also have three charge states, 
but $K^0$ is distinct from its antiparticle, having the strangeness quantum 
number of $+1$.  This was an important finding made by Nakano and Nishijima 
\cite{NakanoNishijima} and Gell-Mann \cite{GellMann}.  
Discovery of antiprotons \cite{antiproton} and antineutrons \cite{antineutron} 
followed in 1955 and 1956, by accelerator experiments at Bevatron in Berkeley. 
The energy of the accelerator had been determined for this purpose.  
A large number of new particles have been discovered since then, but we have 
no reason to doubt that each particle is accompanied with its antiparticle.  

The reasoning of Dirac to predict (in one year before the discovery) 
the existence of positron was the well-known Dirac sea, an assumption that 
the vacuum is the state in which all negative energy states are already 
filled by the particle (electron).  Pauli principle dictates that any 
electron with positive energy cannot drop into a negative energy state 
by emitting a photon etc., so that the existence of negative energy states 
does not affect the stability of an electron state e.g.\ at rest.  
Existence of the negative energy states can be perceived when a negative 
energy electron in the vacuum is excited up to a positive energy state.  
The absence of a negative energy state, a hole in the vacuum, is seen as 
a particle having an opposite charge to the electron.  
This is the antiparticle, the positron.  

This picture is totally asymmetrical between particles and antiparticles, 
and can be applied to fermions only.  Charge of the negative energy 
electrons, total of which is infinite, is ignored.  
In relativistic quantum field theories, 
however, it is  not necessary to evoke the Dirac sea.  
Existence of antiparticle can be shown from general principles for both 
fermions and bosons.  Moreover, the theory is symmetrical between particles 
and antiparticles.  The question that which is particle and which is 
antiparticle makes no sense. 

Space-time picture of particle trajectories {\it \`a la\/} Feynman gives 
an intuitive reasoning for the existence of antiparticles.  It is easier to 
examine an example.  Let us take the nuclear force between a proton and a 
neutron caused by the exchange of a charged pion.  The proton emits $\pi^+$ 
and becomes a neutron, then the neutron absorbs the $\pi^+$ and becomes a 
proton.  The propagation of pion may be described by Feynman propagator, 
which does not vanish outside the light cone, though falling off 
exponentially.  In a different coordinate frame, the timing of the emission 
and absorption becomes the opposite, the $\pi^+$ travels backward in time.  
According to the interpretation of Stueckelberg \cite{Stueckelberg} 
and Feynman \cite{FeynmanPositron}, this should 
rather be described as $\pi^-$ emission by the neutron, followed by its 
absorption by the proton.  In this manner the particle traveling backward 
in time represents the antiparticle.

\section{CPT theorem}

In relativistic quantum field theories, the existence of antiparticle is 
ensured by the CPT theorem, which states that CPT is a symmetry of any 
relativistic quantum field theory under very general assumptions.  
There are two kinds of proofs of the CPT theorem. 
L\"uders \cite{Luders1, Luders2} and Pauli \cite{Pauli} gave constructing 
proof of the theorem, which essentially examines all possible forms of 
relativistic local field theory defined by a Lagrangian density, consisting 
of fields at the same spacetime point and being hermitian, and find they are 
indeed invariant under the combination of C, P, and T transformations.  
It is remarkable that their works were done before 
the suggestion \cite{LeeYang} and discovery \cite{Wu} of parity violation 
in weak interactions, not to mention of CP violation \cite{Christenson}.  

Central to their derivation was the transformation properties of 
the spin-1/2 fields.  
For fields of spin-1/2 particle, CPT transformation of $\psi(x)$ 
(in four component notation) is given by
\begin{equation}
{\cal CPT}\psi(\vec x,t)({\cal CPT})^{-1} = i\gamma_5^* \psi^*(-\vec x,-t)\;.
\end{equation}
Consider any bilinear form consisting of two spin-1/2 fields
\begin{equation}
f \bar\psi_1(x)\Gamma\psi_2(x)
\end{equation}
where $f$ is a complex number and $\Gamma$ is one of the $4\times4$ matrices 
$$\Gamma= \{ 1, \gamma_5, \gamma^\mu, \gamma^\mu\gamma_5, \sigma^{\mu\nu} \} 
\;.$$
It transforms under CPT as
\begin{equation}
\label{CPToffermionbilinear}
{\cal CPT}f \bar\psi_1(x)\Gamma\psi_2(x)({\cal CPT})^{-1} 
= \pm f^* \bigl(\bar\psi_1(-x)\Gamma\psi_2(-x)\bigr)^\dagger \;,
\end{equation}
where Fermi statistics is assumed.  The sign in (\ref{CPToffermionbilinear}) 
counts the number of gamma matrices in $\Gamma$, and is 
$-1$ for $\gamma^\mu$ and $\gamma^\mu\gamma_5$, $+1$ for others.  
Thus the bilinear transforms to its hermitian conjugate up to a possible 
minus sign when it is a Lorentz vector.

This matches with CPT transformation of scalar and vector fields 
\begin{align}
{\cal CPT}\varphi(x)({\cal CPT})^{-1} = \varphi^*(-x)\;,\\
{\cal CPT}A_\mu(x)({\cal CPT})^{-1} = -A_\mu^*(-x)\;.
\end{align}
If we note that the derivative $\partial/\partial x^\mu$ changes sign 
under CPT (since $x\to -x$), it is easy to 
convince oneself that any Lorentz scalar formed from fields at the same 
spacetime point $x$ is transformed by CPT to its hermitian conjugate at $-x$.  
As the Lagrangian density ${\cal L}$ is hermitian as a whole, it transforms 
as ${\cal L}(x)\to{\cal L}(-x)$, and its integral over spacetime, the action, 
is invariant under CPT.  

Pauli \cite{Pauli} also extended these considerations to fields with general 
values of spin.  It is convenient to use fields in irreducible 
representations of Lorentz group.  Using the local isomorphism 
$\rm SO(4) \sim SO(3)\otimes SO(3) \sim SU(2)\otimes SU(2)$, the 
representations can be characterized by two SU(2) `spin' quantum 
numbers, $J_L$ and $J_R$.  For example, a Dirac field corresponds to the 
representation $(1/2,0)\oplus(0,1/2)$ (which corresponding to the left- 
and right-handed Weyl components), and a vector field corresponds to 
the (1/2, 1/2) representation. 
CPT transformation of a general field $(J_L, J_R)$ is found 
to be $(-1)^{2J_L}$ times the conjugated fields with an extra factor of 
$i$ for fermions.  (The factor can alternatively be taken as $(-1)^{2J_R}$.  
The two possibilities are related by a $2\pi$ rotation.)  
The reader may convince him/herself that this general rule reproduces 
the $\gamma_5$ factor for Dirac fields.  
Using this transformation property, CPT invariance of a general Lagrangian 
density can be shown if anticommutation relations for spin half-integer 
fields are assumed.

Second proof of the CPT theorem is the so-called axiomatic one 
\cite{Jost1,Dyson,Jost2,StreaterWightman,Greenberg}.  
We start from the understanding \cite{Wightman} that 
knowledge of Wightman functions, vacuum expectation values of 
products of the fields, of a field theory 
includes the whole information of the theory.  
Thus we are led to examine the CPT transformation properties of Wightman 
functions.

Lorentz group consists of transformations $x\to \Lambda x$ which leaves 
the Minkowski metric invariant
\begin{equation}
\Lambda g \Lambda^T = g
\end{equation}
in matrix notation.  Lorentz group has four disconnected components, each 
of which contains the identity, P, T, and PT, respectively.  These 
components are characterized by $\det\Lambda=\pm1$ and the sign of 
$\Lambda^0{}_0$ (note that $|\Lambda^0{}_0|\ge1$), being $(++)$, $(-+)$, 
$(--)$, and $(+-)$.  
The principle of relativity is the statement that physics laws have 
the same form in any inertial frames, but we learned in 1957 \cite{Wu} that 
the statement is restricted to frames related with each other by 
continuously connected transformations, namely, only the one of the four 
components containing identity.  

Now an important observation is that the identity component and 
the PT component can be connected continuously if we extend the 
Lorentz group to allow complex parameters.  In fact, the boost in 
$x$ direction
\begin{equation}
\begin{pmatrix} t'\\ x' \end{pmatrix}
=\begin{pmatrix} \cosh\eta & \sinh\eta\\ 
\sinh\eta & \cosh\eta \end{pmatrix}
\begin{pmatrix} t\\ x \end{pmatrix}
\end{equation}
leaves $g$ invariant even when $\eta$ takes complex values.  Putting 
pure imaginary $\eta=i\chi$, one finds
\begin{equation}
\begin{pmatrix} t'\\ x' \end{pmatrix}
=\begin{pmatrix} \cos\chi & i\sin\chi\\ 
i\sin\chi & \cos\chi \end{pmatrix}
\begin{pmatrix} t\\ x \end{pmatrix}\;.
\end{equation}
Taking $\chi=\pi$, it reduces to $(t', x') = (-t, -x)$.  
Combining this tranformation with a $\pi$ rotation in the $yz$ 
plane, one arrives at PT, total reflection $x^\mu\to -x^\mu$. 

For this transformation to be useful, the coordinate space needs to 
be extended to complex values, and analyticity of Wightman functions 
needs to be studied.  
For simplicity of presentation and to underline essential points, 
we will look at a two-point function of the form
\begin{equation}
\label{twopointfunction}
\Delta(x_1, x_2) = \langle0| \varphi_1(x_1) \varphi_2^*(x_2) |0\rangle
\end{equation}
where the fields are taken to be scalar fields for the moment.  
(Here we use complex fields and distinguish $\varphi_1$ and $\varphi_2$ 
to highlight the role of charge conjugation. 
The reader may claim that it is nonzero only for $\varphi_1=\varphi_2$, 
but in general $n$-point functions all fields can be different from one 
another.  Even for two point functions, we may regard 
(\ref{twopointfunction}) as something like 
$\langle0| \nu_e(x_1) \bar\nu_\mu(x_2) |0\rangle$ ignoring spin.) 
$\Delta(x_1, x_2)$ turns out to be a function of $x=x_1-x_2$ only, 
thanks to translation invariance 
(note $\varphi(x)= e^{i{\cal P}\cdot x} \varphi(0) e^{-i{\cal P}\cdot x}$. 
The vacuum, having $P^\mu=0$, is taken to be translation invariant): 
\begin{equation}
\Delta(x_1, x_2)=\Delta(x_1-x_2,0) 
= \langle0| \varphi_1(x) \varphi_2^*(0) |0\rangle \;,
\end{equation}
which we will write simply as $\Delta(x)$. 
When $x^\mu$ is extended to complex values $z^\mu$, its analyticity 
property can be seen to be related to the spectrum of the physical states.  
Inserting a complete set of states, it may be expressed as
\begin{equation}
\begin{split}
\Delta(z) &= \langle0| \varphi_1(z) \varphi_2^*(0) |0\rangle
= {\textstyle\sum\limits_n} \langle0| \varphi_1(z) |n\rangle 
\langle n|\varphi_2^*(0) |0\rangle \\
&= {\textstyle\sum\limits_n} \langle0| \varphi_1(0) |n\rangle e^{-i p_n\cdot z}
\langle n|\varphi_2^*(0) |0\rangle \\
&= \int_{m^2}^\infty \!\! dM^2\, \rho_{12}(M^2) 
\int{d^3p\over(2\pi)^3 2\sqrt{\vec p^2+M^2}}\,e^{-ip\cdot z} \;.
\end{split}
\end{equation}
The four momentum $p^\mu$ in the last line satisfies physical 
condition $p^0 = \sqrt{\vec p^2+M^2} > |\vec p|\ge0$.  
Writing $z^\mu=x^\mu+iy^\mu$, the integral on three momenta can be 
seen to converge for $y^0<0$ and $|y^0|>|\vec y|$, {\it i.e.} 
when $y^\mu$, imaginary part of $z^\mu$, is inside the backward light cone.  
This region (with $x^\mu$ arbitrary) is traditionally called 
the `backward tube.'

The vacuum expectation value $\langle0| \varphi_1(z) \varphi_2^*(0) |0\rangle$ 
itself is a Lorentz invariant quantity, so it is expected to be a function 
of the only scalar $z^2$.  In the above, the domain of analyticity is seen 
to be the backward tube.  It is shown that this region can be extended 
to the union of the region $z \to \Lambda_c z$ by analytic continuation, 
where $z$ is inside the backward tube 
and $\Lambda_c$ is a complex Lorentz transformation \cite{HallWightman}.  
This extended region of analyticity (`the extended tube') contains 
points where $z^2$ is real (`Jost points').  
This is realized when $x\cdot y=0$ (then $z^2=x^2-y^2$), which implies 
that $x^\mu$ is a spacelike four vector.  

This result allows to transform the Wightman functions under 
complex Lorentz transformation keeping their analyticity.  
Using Lorentz invariance of the vacuum and Lorentz transformation 
properties of the fields 
${\cal U}(\Lambda) \varphi(z) {\cal U}^\dagger(\Lambda) 
= S(\Lambda)^{-1} \varphi(\Lambda z)$ (for scalar fields $S=1$), 
and taking ${\cal U}$ as the total reflection, the relation 
\begin{equation}
\label{PTrelation}
\langle0| \varphi_1(z) \varphi_2^*(0) |0\rangle
= \langle0| \varphi_1(-z) \varphi_2^*(0) |0\rangle
\end{equation}
is obtained.  At first glance this looks like a PT invariance 
relation of Wightman functions, but it is not necessarily so.  
Especially when $x^2$ is timelike, the region of analyticity of the two 
sides has no common region.  The right-hand side of (\ref{PTrelation}) 
in fact converges when $y^\mu$ is inside the {\it forward\/} light 
cone.  This in practice means that there can be discontinuity in the 
limit $y^\mu\to\pm0$.  
This is indeed the case for free fields, 
where the discontinuity is given by Bessel function.  

Now we assume that for spacelike separation the fields commute 
(anticommute for fermions), not necessarily as operators but 
at least inside vacuum expectation values (this is called `weak 
local commutativity').  Then we have
\begin{equation}
\label{almostCPT}
\begin{split}
\langle0| \varphi_1(z) \varphi_2^*(0) |0\rangle
&= \langle0| \varphi_1(-z) \varphi_2^*(0) |0\rangle \\
&= \langle0| \varphi_2^*(0) \varphi_1(-z) |0\rangle \;.
\end{split}
\end{equation}
Using translation invariance, the second row may also be written as
$\langle0| \varphi_2^*(z) \varphi_1(0) |0\rangle$, so we have
\begin{equation}
\label{PTplusLWC}
\langle0| \varphi_1(z) \varphi_2^*(0) |0\rangle
= \langle0| \varphi_2^*(z) \varphi_1(0) |0\rangle \;.
\end{equation}
Though the equality holds for spacelike $x^\mu$, 
the domain of convergence for the two sides is now common, 
both in backward tube.  
The equality can then be extended to the whole region of $z$ by 
analytic continuation.  

One more twist is needed.  
When we express the right-hand side of (\ref{PTplusLWC}) in terms of 
$t'=-t$, it results in the wrong sign of frequency.  (The canonical 
commutation relation also has a wrong sign with $\vec x'=-\vec x$ and 
$\vec p'=\vec p$).  As was realized by Wigner, this term 
should rather be viewed as
\begin{equation}
\langle0| \varphi_2^*(z) \varphi_1(0) |0\rangle
= \langle0| \varphi_1^*(0) \varphi_2(z) |0\rangle^*
\end{equation}
for consistent interpretation of time reversal in quantum theory 
(antiunitarity of transformations including time reversal).  
So we finally arrives at the relation 
(translating back in $z$ to the form in RHS of (\ref{almostCPT}))
\begin{equation}
\langle0| \varphi_1(x) \varphi_2^*(0) |0\rangle
= \langle0| \varphi_1^*(-x) \varphi_2(0) |0\rangle^*
\end{equation}
or with the original coordinates
\begin{equation}
\langle0| \varphi_1(x_1) \varphi_2^*(x_2) |0\rangle
= \langle0| \varphi_1^*(-x_1) \varphi_2(-x_2) |0\rangle^*
\end{equation}
which is the CPT relation for Wightman functions, recalling that 
the field conjugation corresponds to the exchange of particles 
and antiparticles.  

If the fields have nonzero spin, Lorentz transformation involves a nontrivial 
matrix $S$ which mixes spin components.  For total reflection, $S$ just 
reduces to a sign factor $(-1)^{2J_L}$.  In addition for fermion fields, 
exchanging them for spacelike distances provides an extra minus sign.  
So the conclusion for scalar fields can be applied to fields with 
arbitrary spin with a possible minus sign.  

It is possible to extend these arguments to vacuum expectation values of 
$n$ fields in a rather straitforward way.  
We do not discuss the details except mentioning that 
theory of multivariable complex functions need to be invoked.  

To summarize, CPT symmetry can be proved in general field theories, 
having Lorentz invariance, physical spectrum, and local commutativity 
consistent with causality.  

Experimental tests of CPT invariance have been performed in a variety 
of observables \cite{ConservPDG}.  The most precise verification 
comes from the difference of $K^0$ and $\bar K^0$ masses 
\cite{Angelopoulos,KaonCPTPDG}, which is less than $10^{-18}$ of 
the mass itself (almost the Planck scale if the breaking effect is linear).  
Magnetic moments of positive and negative muons are equal 
at the level of $10^{-8}$ \cite{Bennett}.

\section{Relativity and set of states}

It is instructive to find out how Poincar\'e invariance constrains 
possible set of particle states (not fields!).  
Consider a particle at rest.  In a different (boosted) coordinate frame, 
the particle has a finite momentum.  
Principle of relativity then implies in turn that states with finite momenta 
have to exist in the original frame.  Thus the whole set of particle states 
with arbitrary three-momenta should be there.  Mathematically, the set 
of states has to form a (irreducible unitary) representation of Poincar\'e 
algebra.  Now working in the rest frame, spacial rotations do not change 
the momentum and the state remains at rest.  How the state transforms under 
rotations is prescribed by representation of SO(3), the group of space 
rotation, which is angular momentum states designated by the spin quantum 
number of the particle.  If the spin of the particle is $S$, the number of 
states is $(2S+1)$ times the spacial momentum degrees of freedom ($\infty^3$).  

The situation is different if the particle has no mass.  The argument 
above is not applicable since massless states have no rest frame. It can 
be easily seen that states with any momentum (except 0) can be connected 
by rotation and boost, forming a set of states.  Now instead of the 
state at rest, we may select a state with one particular momentum as 
the standard and examine transformations which do not change the momentum.  
Spacial rotations generally change the momentum of states, 
but rotation along the momentum vector leaves the momentum unchanged.  
Angular momentum along the momentum vector can thus be well defined.  
We call it `helicity' of the particle, symbolically written as
\begin{equation}
\lambda = {\vec S\cdot\vec p\over |\vec p|} \;.
\end{equation}
It can be shown that helicity is invariant under any Lorentz transformation 
including boost and rotation.  (For massive states, we can define helicity 
of a particle similarly, but it is only invariant under spacial rotation, 
and boost along the momentum direction as long as the momentum direction 
is not reversed. It is not invariant under boost in the transverse directions.)
The states with a single particular value of helicity close within themselves 
under Lorentz transformation.  The number of degrees of freedom for 
massless states is $1\times\infty^3$, as far as Poincare invariance is 
concerned.  

CPT invariance poses additional constraints on massless states.  
Considering the transformation properties of momentum and angular momentum 
under CPT, one can convince that helicity changes sign under CPT.  
Thus transformation of a particle state with helicity $\lambda$ is 
an antiparticle state with helicity $-\lambda$.  The two states with 
opposite helicities should exist, except for the case $\lambda=0$.  
The number of degrees of freedom becomes $2\times\infty^3$ with this 
CPT requirement.

\section{Charge conjugation and self-conjugate particles}

Charge conjugation is the operation of exchanging particles and antiparticles.  In terms of fields, charge conjugation is deeply related to hermitian 
conjugation.  Generally, (free) fields can be expanded in terms of plane 
waves.  Their positive frequency parts are associated with annihilation 
operators of the particle, and negative frequency parts include creation 
operators of the antiparticle.  When the field is conjugated, the negative 
and positive frequencies are exchanged, and annihilation and creation 
operators are exchaged at the same time.  The hermitian conjugated field 
behaves exactly as the antiparticle field: One can see that the conjugated 
field contains annihilation operators of the antiparticles, in constrast to 
the original field containing annihilation operators of the particle.

This relation is most simply realized for scalar fields.  A real scalar field 
corresponds to scalar particle which is identical to its antiparticle, and 
a complex scalar field corresponds to particle which is different from its 
antiparticle.  Charge conjugation is nothing but hermitian conjugation.  

We note that a complex scalar field can be formed from two real scalar fields 
which constitutes real and imaginary parts of the field.  When this is done, 
the interpretatation of particles and antiparticles changes.  In the 
opposite direction, starting from a complex scalar field, we may 
impose a reality condition 
$\varphi^*(x)=\varphi(x)$ to obtain a real scalar field.

Charge conjugation becomes slightly more complicated for Dirac fields. 
The complex conjugation of a Dirac field $\psi^*(x)$ (by the 
simbol $*$, hermitian conjugation as an operator, but no transposition 
as a four-spinor is meant) can play the role of antiparticle field.  
The problem is that $\psi^*$ transforms differently from $\psi$ under 
Lorentz transformations.  If we impose a reality condition $\psi^*=\psi$ 
to obtain a field of self-conjugate spin-1/2 particles, it may work 
in that particlar coordinate frame, but the condition does not hold 
in different frames.  

It is thus convenient if we can find a linear combination of the four 
components of $\psi^*$ which transforms exactly in the same way as $\psi$.  
This is possible (thanks to the equivalence of all representations of 
gamma matrices in four dimensions), and it is usually written as
\begin{equation}
(\psi^c)_\alpha = C_{\alpha\beta} \bar\psi_\beta
= C_{\alpha\beta} (\gamma_0^T)_{\beta\gamma} \psi^*_\gamma
\end{equation}
where $C$ is a constant unitary $4\times4$ matrix satisfying
\begin{equation}
\label{chargeconjugationmatrix}
C \gamma_\mu^T C^{-1} = -\gamma_\mu\;.
\end{equation}
The field $\psi^c$ is called the charge conjugation of $\psi$.  

We do not quote an explicit form of $C$, as it generally depends 
on the representation of $\gamma_\mu$.  
Gamma matrices can be any set of four matrices satisfying 
$\gamma_\mu\gamma_\nu + \gamma_\nu\gamma_\mu=2g_{\mu\nu}$ and 
appropriate hermiticity properties (for the Lagrangian to be hermitian).  
It is known (from representation theory of finite groups) that 
any set satisfying these relations are unitary equivalent to each other, 
thus corresponding to a different choice of the basis of the four-component 
spinor space.  
The set $\{-\gamma_\mu^T\}$ obviously satisfies the same relations as 
$\{\gamma_\mu\}$, and the unitary transformation connecting the two set is 
nothing but the matrix $C$.  

There exist sets of $\{\gamma_\mu\}$, all of which are purely 
imaginary.  These are called Majorana representation.  For this 
representation we can take $C=-\gamma_0$ ($C\gamma_0^T=1$), 
and then charge conjugation is simply complex conjugation, 
just as for scalar fields.  
Indeed it is easy to see that Dirac equation in this representation can be 
split into its real and imaginary parts.  
Unlike Dirac and chiral representations of the gamma matrices, however, 
transformation property under space rotation is cumbersome for 
Majorana representation. 

The fields $\psi$ and $\psi^c$ transforms in the same way under Lorentz 
transformation and both satisfy Dirac equation of the same form.  So it 
is legitimate to impose a condition
\begin{equation}
\psi^c(x)=\psi(x)
\end{equation}
to obtain a field of self-conjugate spin-1/2 particle.  
This condition is called Majorana condition, and 
the resulting field is named Majorana field.  Expanding $\psi$ in 
plane waves it can be seen that antiparticle and particle are identical 
for Majorana fields.  The physical degrees of freedom becomes half 
of that of Dirac fields.

\section{Spin-1/2 fields and mass}

The physical content of Dirac field has four degrees of freedom 
(putting the momentum aside), spin up and down, particle and antiparticle.  
This is totally appropriate to describe electrons.  
It is however not the minimal spin 1/2 field as have been seen in the 
previous section.  Majorana field is one possibility of reducing the degrees 
of freedom to half, but there is a different way of reduction. 

The matrix $\gamma_5$, having the eigenvalues $\pm1$, may be used to split 
a Dirac field into two parts. 
Explicitly, left- (right-)handed projection can be performed by 
multiplying Dirac spinor by $(1\mp\gamma_5)/2$.  
The result is the left-handed and right-handed Weyl (chiral) fields.  
\begin{equation}
\psi = \psi_L + \psi_R\;,\quad 
\gamma_5\psi_L = -\psi_L\;,\quad 
\gamma_5\psi_R = +\psi_R\;.
\end{equation}
However, projecting the Dirac equation $(i\gamma^\mu\partial_\mu-m)\psi=0$ 
into chiral parts we find
\begin{equation}
i\gamma^\mu\partial_\mu \psi_L = m\psi_R \;,\quad
i\gamma^\mu\partial_\mu \psi_R = m\psi_L \;,
\end{equation}
which means that left- and right-handed parts are inter-related to 
each other.  An exception occurs for $m=0$, in which case the two 
decouple.  It is then possible to retain one chirality of Weyl field 
and discard the other.  

As every particle physicist knows well, the Standard Model of fundamental 
interactions is constructed using Weyl fields as its constituent fields 
representing quarks and leptons.  All these fermions have 
no mass terms in the Lagrangian, and obtain masses only after 
spontaneous $\rm SU(2)\otimes U(1)$ breaking through Yukawa couplings 
to the Higgs field. 

The physical content of Weyl fields is half of that of Dirac fields 
obviously, but 
it is notable that the left-handed Weyl fields contain negative helicity 
particle and {\it positive\/} helicity antiparticle.  
Weyl fields themselves thus break 
parity and charge-conjugation invariance from outset.  
This content is consistent with CPT invariance which is respected by any 
relativistic field theory.  
CPT transformation of a negative helicity particle state is indeed 
a positive helicity antiparticle state.  

One may wonder that imposing Majorana condition and chiral projection 
at the same time may reduce the physical degrees of freedom further.  
It turns out that this is not possible.  Consideration of CPT invariance 
in fact shows that two degrees of freedom is minimal for spin 1/2 
particles.  If a Majorana field $\psi_M$ is projected to chiral components, 
each projection no longer satisfies Majorana condition, and just becomes a 
usual Weyl field.  The left- and right-handed Weyl fields are however 
not independent and related to each other
\begin{equation}
\psi_R=(\psi_L)^c \;.
\end{equation}
(The interchange of left- and right-handed components may be understood if 
one reminds the charge conjugation property of Weyl fields.)  
In turn, starting from a Weyl field $\psi_L$, it is possible to 
construct a Majorana field as
\begin{equation}
\psi_M=\psi_L + (\psi_L)^c \;.
\end{equation}
These considerations imply that Weyl and Majorana fields have the same 
physical content and are equivalent in a sense. 

We saw that Weyl fields admit no mass terms.  In contrast, Majorana fields 
can have mass terms, which may seem contradictory.  There is actually 
no contradiction, as the Majorana mass term enters as
\begin{equation}
i\gamma^\mu\partial_\mu \psi_L = m(\psi_L)^c \;,
\end{equation}
from which it can be seen that conserved charge no longer exists.  
In terms of Weyl fields, the Majorana mass violates the particle number 
conservation.  In the case of neutrinos, lepton number conservation 
is violated if neutrino has a Majorana mass.

Lagrangian density for free Majorana field $\psi_M$ has the following form
\begin{equation}
{\cal L} = {1\over2}\bigl(\bar\psi_M i\gamma^\mu \partial_\mu \psi_M 
- m \bar\psi_M\psi_M\bigr) \;.
\end{equation}
This is the same form as that for Dirac field up to the overall 
factor of 1/2.  This factor is essentially the same as the difference 
between the Lagrangian of real and complex scalar fields.  
Rewriting the Lagrangian in terms of corresponding left-handed Weyl field 
$\psi_L$, we find up to total divergence
\begin{equation}
{\cal L} = \bar\psi_L i\gamma^\mu \partial_\mu \psi_L
- {1\over2} m \bigl(\bar\psi_L (\psi_L)^c + (\bar\psi_L)^c \psi_L \bigr) \;.
\end{equation}
The mass term may be rewritten as
\begin{equation}
\begin{split}
\bar\psi_L (\psi_L)^c + (\bar\psi_L)^c \psi_L 
&= \bar\psi_L C \bar\psi_L^T  - \psi_L^T C^\dagger \psi_L\\
&= C_{\alpha\beta} (\bar\psi_L)_\alpha (\bar\psi_L)_\beta 
+ C_{\alpha\beta}^* (\psi_L)_\alpha (\psi_L)_\beta \;,
\end{split}
\end{equation}
which explicitly shows that the Majorana mass violates fermion 
number of the Weyl field.  Notice that the mass term is compatible 
with Fermi statistics, as $C$ is an antisymmetric matrix. 

In the following we use shortcut notation
\begin{equation}
\bar\psi_L \cdot \bar\psi_L + \psi_L\cdot\psi_L
\end{equation}
for this Majorana mass term.  

In terms of the Majorana field $\psi_M$, the mass term $\bar\psi_M\psi_M$ 
may be rewritten as
\begin{equation}
\bar\psi_M\psi_M = -\psi_M^T C^{-1} \psi_M = \bar\psi_M C \bar\psi_M^T \;.
\end{equation}
From this expression also it can be seen that antisymmetry of $C$ and 
Fermi statistics go together.  Extending this argument to general Majorana 
bilinears $\bar\psi_M \Gamma \psi_M$, symmetry of the product $\Gamma C$ 
under transposition turns out to be essential.  If $\Gamma C$ is a symmetric 
matrix, $\bar\psi_M \Gamma \psi_M$ {\it vanishes\/} automatically.  
This occurs for $\Gamma=\gamma^\mu$ and $\sigma^{\mu\nu}$.  
Absence of the vector bilinear is nothing but the property that Majorana 
fermions cannot have any charges.  Similarly, Majorana fermions cannot have 
magnetic or electric dipole mements.  Other bilinears $\Gamma=1$ (mass term), 
$\gamma_5$, $\gamma^\mu\gamma_5$ are nonzero.  
Coupling of Majorana fermions to gauge bosons thus is not in general 
forbidden but is restricted to axial vector coupling.  
For bilinears of different Majorana fields $\bar\psi_{M1}\Gamma\psi_{M2}$, 
these consideration do not apply and vector transition coupling is allowed.

\section{Neutrino masses}

It is often stated that neutrino masses indicate the existence of 
physics beyond the Standard Model.  
Rather, the Standard Model was deliberately constructed to allow no 
neutrino masses, 
by not introducing right-handed neutrinos in the model. 
This came from the experimental observation that 
the neutrinos from weak decays are always 
left-handed \cite{Goldhaber}.  

There is no evidence that right-handed neutrinos exist, but their 
expected quantum numbers 
imply that they are singlet of the 
Standard Model gauge group
and have no gauge interactions.  
Thus right-handed neutrinos, even if they exist, are hard to 
produce in particle reactions we can handle. 

With right-handed neutrinos, the fermion content of the Standard Model 
becomes symmetric between quarks and leptons, and neutrino masses arise 
just in the same way as the charge-$2/3$ quark masses from Yukawa couplings 
to the Higgs field.  The size of the neutrino masses is expected to be 
of the same order of the corresponding charged lepton, which strongly 
disagrees from the observation.  From direct measurements of tritium 
beta decay, the electron neutrino mass is less than 0.8 eV \cite{Aker}.  
From early cosmology data including CMB constrain the sum of the neutrino 
masses to be less than around 0.1 eV \cite{Planck,NeutrinoCosmology}.  
This mass range is more than a million times smaller than the mass of the 
electron, the lightest charged particle.

A clue to understand the smallness of the neutrino masses may be the fact 
that neutrinos have no electric charge.  Unlike charged quarks and leptons, 
neutrinos can possibly be Majorana fermions.  Although lepton number 
distinguish neutrinos from antineutrinos, lepton number conservation 
is based on a global symmetry and is not on a firm ground compared 
to gauge symmetries.  Introduction of neutrino Majorana mass terms 
to the Standard Model can in fact be done in several ways.  

The left-handed neutrino forms an SU(2) doublet with the corresponding 
charged lepton. 
Thus the Majorana mass term for the `left-handed' neutrino 
$\nu\cdot\nu + {\rm h.c.}$ behaves as $I=I_3=-Y=1$ in the Standard Model 
gauge group and is not invariant.  
If there is an extra triplet Higgs field $\chi$ with $I=Y=1$, gauge invariant 
Yukawa coupling can be written in the form 
(for simplicity we do not write generation degrees of freedom)
\begin{equation}
\label{tripletHiggscoupling}
{\cal L} = -h \ell_L^T\cdot i\tau_2\tau_a \ell_L\,\chi_a + {\rm h.c.} \;,
\end{equation}
where $\ell_L$ is the doublet left-handed lepton field, 
$h$ is the triplet Yukawa coupling, and $\tau_a$ is the gauge SU(2) Pauli 
matrix.  ($i\tau_2$ is not physically the second component of Pauli matrix, 
but the `charge conjugation' matrix of internal SU(2), which makes 
an SU(2) doublet $\varphi$ and its complex conjugate $\varphi^*$ 
transform exactly in the same way under SU(2) transformation.  
Thus $\varphi^* i\tau_2\tau_a\varphi$ is the spin-1 combination formed 
by the product of two spin 1/2.  
So it may be better to use a different notation such as $\epsilon$ instead 
of $i\tau_2$.) 
When the neutral component of the triplet Higgs field develops 
a vacuum expectation value, the interaction terms (\ref{tripletHiggscoupling}) 
produces a neutrino Majorana mass of the order $h\langle \chi\rangle$.  
Smallness of the neutrino masses may be due to the smallness of the Yukawa 
coupling $h$ and/or the smallness of the triplet vev.

This scenario requires the introduction of a new scalar field, but 
one may notice that a triplet scalar operator can be formed from the usual 
doublet Higgs field as
\begin{equation}
\varphi^T (i\tau_2\tau_a) \varphi
\end{equation}
without introducing new fields.  An interaction term 
\begin{equation}
\label{dimensionfive}
{\cal L} = -{1\over\Lambda}\,\varphi^T (i\tau_2\tau_a) \varphi \;
\ell_L^T\cdot (i\tau_2\tau_a) \ell_L + {\rm h.c.}
\end{equation}
may be written down as a gauge invariant interaction.  This operator 
has dimension five and is not renormalizable, as reflected in the 
negative dimension factor $1/\Lambda$ in front of the operator.  
When ${\rm SU(2) \otimes U(1)}_Y$ is broken in the usual way, the 
interaction (\ref{dimensionfive}) gives a Majorana mass to the neutrino 
\begin{equation}
m_\nu \sim {v^2\over\Lambda} \;,
\end{equation}
where $v$ is the usual Higgs vev ($\sim 250$ GeV).
If the mass scale $\Lambda$ is very large, the resulting neutrino 
Majorana mass becomes tiny.  

It is possible to make up a renormalizable `UV completion' of this scenario by 
introducing right-handed neutrinos.  Being gauge singlets, right-handed 
neutrinos can have gauge invariant Majorana mass terms in the Lagrangian.  
In addition, Yukawa couplings between left-handed lepton doublet 
and the right-handed neutrino can be included to give the usual 
Dirac mass.  The relevant terms in the Lagrangian are
\begin{equation}
{\cal L} = -{1\over2}M_R \nu_R\cdot\nu_R 
- f_\nu \varphi^\dagger \bar\nu_R \ell_L + {\rm h.c.} \;.
\end{equation}
It is convenient to 
rewrite the right-handed neutrino fields using their conjugate left-handed 
antineutrino fields $N_L=(\nu_R)^c$, so  
\begin{equation}
{\cal L} = -{1\over2}M_R N_L\cdot N_L
- f_\nu \varphi^\dagger N_L \cdot \ell_L + {\rm h.c.} \;.
\end{equation}
%

The neutrino mass terms after ${\rm SU(2) \otimes U(1)}_Y$ breaking 
form a $2\times2$ matrix
\begin{equation}
{\cal L}_{\rm mass} = - {1\over2}
\begin{pmatrix} \nu_L & N_L\end{pmatrix}
\cdot
\begin{pmatrix} 0 & f_\nu v/\sqrt2\\ f_\nu v/\sqrt2 & M_R \end{pmatrix}
\begin{pmatrix} \nu_L \\ N_L \end{pmatrix}
 + {\rm h.c.} \;.
\end{equation}
Notice that the right-handed Majorana mass $M_R$ is independent from 
the ${\rm SU(2) \otimes U(1)}_Y$ breaking scale (the Higgs mass term) and 
can be very large, while the Yukawa coupling $f_\nu$ naturally lies in 
the same range as the other Yukawa couplings.  
If so, diagonalizing the matrix gives two mass eigenstates with masses 
$\sim M_R$ and $m_\nu \sim(f_\nu v)^2/M_R$.  The mixing of the two 
states is of the order $\sim f_\nu v/M_R$, so the lighter state essentially 
consists of the left-handed (doublet) neutrino.  
This scenario \cite{GMRS,Yanagida} is called the `seesaw mechanism.' 
If one take $m_\nu\sim 0.1$ eV and assume the Dirac mass term of 
1 MeV--100 GeV, $M_R$ ranges from $10^4$ to $10^{14}$ GeV.

The Majorana mass term violates lepton number conservation.  This effect 
is essentially invisible at high energies, where the mass can be neglected.  
The most promising process to detect lepton number violation is 
neutrinoless double beta decay.  

It is known that some nucleus stable against beta decay can decay to 
the second neighboring nucleus by emitting two electrons 
and two neutrinos: 
\begin{equation}
(Z,A) \to (Z+2,A) + 2 e^- + 2\bar\nu_e \;,
\end{equation}
where $Z$ is the atomic number (number of protons) of the nucleus and 
$A$ is the mass number (number of protons and neutrons) of the nucleus.  
This is the second-order process of beta decay and has a very long lifetime.  
Around ten kinds of nuclei have been observed to decay in this way, 
and the measured lifetimes are in the range of $10^{18}$--$10^{24}$ years 
\cite{Saakyan}.  

If neutrinos have Majorana masses and lepton number is violated, 
double beta decay without emitting neutrinos becomes possible. 
\begin{equation}
(Z,A) \to (Z+2,A) + 2 e^- \;.
\end{equation}
The matrix element of the process is proportional to a combination of 
neutrino Majorana masses and neutrino mixing angles/phases.  
The sum of the energy of the two electrons is monochromatic, 
in contrast to that in the usual double beta decays.  
Searches for neutrinoless double beta decay have been done by a number 
of experiments.  Currently the best limit of $2.3\times10^{26}$ years 
for this type of decay is obtained for ${}^{136}{\rm Xe}$ by KamLAND-Zen 
Collaboration \cite{KamlandZen}.  This limit corresponds to the limit 
on the `effective' Majorana mass of 36--156 meV, depending on the evaluation 
uncertainty of the nuclear matrix element.

\section{Baryon number of the universe and leptogenesis}

We are made of matter, nucleons and electrons.  Not only ourselves, 
we have so far observed no trace of antimatter existing in the universe, 
except for tiny amounts in cosmic rays etc., which can be understood to be 
products from high-energy collisions of normal matter.  
In the very early universe with extremely high temperature ($T\gg 1$ GeV), 
however, it is expected that matter and antimatter existed essentially 
in equal amount. For example, high energy photons can be converted to 
any particle-antiparticle pair when energetically possible.  
When the temperature was lowered, matter and antimatter annihilated with 
each other and a very small amount of matter should have been left.  
The amount of matter can be calculated from the measurement of cosmological 
parameters using cosmic microwave background (CMB) \cite{Planck} and others. 
Comparing it with the known amount of the 2.7 K CMB photons, 
we find the baryon-to-photon ratio 
\begin{equation}
n_B/n_\gamma \sim 0.6 \times 10^{-9} \;.
\end{equation}
At the very early universe baryons and antibaryons have been in thermal 
equilibrium with other particles including photons.  Their number density 
is of the same order as photons.  The asymmetry between baryons and 
antibaryons in the early universe was therefore of the order of $10^{-9}$.  

It has been known that three ingredients are needed to produce asymmetry 
between baryons and antibaryons \cite{Sakharov}: 
(1) baryon number nonconserving process; 
(2) violation of CP (and C) invariance; 
(3) deviation from thermal equilibrium.  

We have no evidence for instability of protons (baryon number 
nonconservation).  Indeed there is no baryon-number violating interactions 
in the Standard Model Lagrangian.  
It was shown \cite{tHooft}, however, that because only left-handed 
fermions couple to the SU(2) $W$ boson (the theory is chiral), quantum 
effect violates conservation of baryon number current.  This is the 
well-known chiral anomaly.  The effect becomes physical only when it 
is coupled with nonpertubative gauge field configuration, and is totally 
negligible at our energy scale.  The effect can be substantial when 
the temperature is above the electroweak scale, and this should be taken 
into account in tracing the history of the early universe \cite{Kuzmin}. 
In addition to baryon number $B$, lepton number $L$ is also violated.  
It should be stressed that the difference $B-L$  is free from anomaly 
and is conserved.

In the Standard Model, CP violation arises from the complex phase of the Yukawa 
couplings, as was shown by Kobayashi and Maskawa \cite{KobayashiMaskawa}.  
Unfortunately its effect on baryon number generation is many orders of 
magnitude smaller than needed to explain the baryon asymmetry 
\cite{Shaposhnikov}.  

A scenario called leptogenesis \cite{FukugitaYanagida} is an attractive 
alternative to derive the baryon asymmetry.  It proceeds in two steps. 
First, asymmetry in the lepton number is created at a very early stage 
of the universe ($T\gg 1$ TeV), when ${\rm SU(2)\otimes U(1)}_Y$ is 
unbroken.  Then, when the temperature of the universe drops to 
around the electroweak scale, transition from the symmetric universe to 
the spontaneously broken state occurs (electroweak phase transition, 
the Higgs field develops a vacuum expactation value).  During the 
transition, a part of the lepton number asymmetry can be transferred 
to baryon number asymmetry when the universe is temporarily out of 
thermal equilibrium.  

The first stage generation of net lepton number can be related to 
right-handed Majorana neutrinos with CP violating interaction.  
The right-handed neutrino (denoted $N$) can decay into the modes 
$N\to \nu_L H$ and $N\to \bar\nu_R H$, because the $N$ Majorana mass 
itself violates lepton number conservation.  If the Yukawa couplings 
contain CP violation, the branching ratios of these two modes can be unequal.  
At very high temperatures, however, reverse reactions (formation of $N$) 
also proceeds at the same rate 
and no net amount of lepton number is produced.  
Departure from this state of thermal equilibrium can be realized 
when the temperature drops to slightly below the $N$ mass.  Because the 
typical energy of particles in the universe becomes insufficient to 
produce $N$, the reverse reaction is suppressed, and lepton number 
can be generated by the asymmetric decay processes.  
With adjusting parameters in the model, this scenario can explain 
the amount of baryon asymmetry \cite{Davidson}, as generation mixing 
of neutrinos is much larger than that of quarks.

\section{Grand Unified Theories}

Grand Unified Theory (GUT) is a gauge theory with a large gauge group 
which contains the three gauge groups of the Standard Model.  Moreover, 
a GUT typically unifies quarks and leptons, putting them in same gauge 
multiplets. The simplest GUT is based on the group SU(5) 
\cite{GeorgiGlashow}.  One generation of quarks and leptons, 
which comprise five irreducible representations 
of ${\rm SU(3)\otimes SU(2)\otimes U(1)}_Y$, are merged into two SU(5) 
representations $5^*$ and 10:
($5^*$ is the conjugate representation of the basic 5 representation of SU(5), 
and 10 is the antisymmetric second-rank tensor representation contained in 
$5\otimes5$.)  
\begin{equation}
\label{SU5fermions}
5^*_L = (d^c, \ell)_L\;,\qquad
10_L = (q, u^c, e^c)_L
\end{equation}
where all the Weyl fields are taken to be left-handed.  
It is not just the quarks and leptons are in the same representations, but 
quarks and antiquarks are in the same 10 representation.  
This shows that baryon number and lepton number cannot 
be defined in the GUT lagrangian.  Only after the breaking of the SU(5) 
gauge symmetry it becomes possible to distinguish quarks from leptons, 
and also quarks from antiquarks.  

The SU(5) gauge field can be put in a traceless $5\times5$ hermitian matrix.  
The first diagonal $3\times3$ part corresponds to color SU(3), 
the remaining diagonal $2\times2$ weak SU(2).  The weak hypercharge 
${\rm U}(1)_Y$, which commutes with SU(3) and SU(2), corresponds to 
a diagonal matrix proportional to $(2, 2, 2, -3, -3)$.  

Mathematically, this expresses that 24-dimensional SU(5) adjoint 
representation may be decomposed into representations of its subgroup 
${\rm SU(3)\otimes SU(2)\otimes U(1)}_Y$ 
as follows:
\begin{equation}
24 = (8,1)_0 \oplus (1,3)_0 \oplus (1,1)_0 
\oplus (3,2)_{-5/6} \oplus (3^*,2)_{5/6} \;,
\end{equation}
where the right-hand side denotes the representation in the form 
${\rm (SU(3), SU(2))}_Y$. The first three corresponds to gluons and 
${\rm SU(2)}$ and ${\rm U(1)}_Y$ gauge bosons.  
We will touch on the remaining 12 states shortly.  
Likewise, decompositions of fermion representations are (as can 
be seen from (\ref{SU5fermions}))
\begin{align}
5^* &= (3^*,1)_{1/3} \oplus (1,2)_{-1/2}\\
10 &= (3,2)_{1/6} \oplus (3^*,1)_{-2/3} \oplus (1,1)_1
\end{align}

The breaking of SU(5) to ${\rm SU(3)\otimes SU(2)\otimes U(1)}_Y$ 
can be attained if a scalar field in the adjoint representation 24 
develops a nontrivial vacuum.  Minimization of a quartic 
scalar potential gives at most two eigenvalues for the vacuum 
expectation value, and one possibility is the form proportional to $Y$, 
resulting in remaining symmetries of the Standard Model gauge group.  

After the GUT symmetry breaking 
${\rm SU(5)\to SU(3)\otimes SU(2)\otimes U(1)}_Y$, 
the extra gauge bosons not contained in the Standard Model gauge group 
develop masses around the GUT scale.  The infrared theory containing only 
`massless' particles (compared to the GUT scale) is the Standard Model, 
In this stage, separate conservation laws of baryon and lepton numbers 
emerge as infrared symmetries 
(barring nonperturbative effect discussed in the previous section).  
Couplings of the extra gauge bosons, however, do not conserve baryon 
and lepton numbers and lead to proton instability.  
Experimental nonobservation of proton decays \cite{SuperKND} 
constrains the GUT scale to be at least as large as $10^{16}$ GeV, 
which coincides with the expectation from the renormalization group 
evolution of the three gauge couplings \cite{GeorgiQuinnWeinberg} 
(better agreement with supersymmetry).

No SU(5) invariant mass terms are allowed for the fermions, 
since no combinations of 
$5^*\otimes 10$, $10\otimes 10$ or $5^*\otimes 5^*$ contain singlets 
(in contrast to $5^*\otimes 5$ and $10\otimes 10^*$ appearing 
in the kinetic terms).  
This should be the case since SU(5) contains the Standard Model gauge group.  
Otherwise the fermions could have gained masses of the order of the GUT scale. 
The Higgs field in the Standard Model can most simply be put in the basic 5 
representation of SU(5).  The Yukawa couplings to generate quark and 
lepton masses are in the invariant combinations 
\begin{equation} 
5^*_L 10_L 5_H^* \;,\qquad
10_L 10_L 5_H \;.
\end{equation}
The former generates down quark and charged lepton masses, and the 
latter is responsible for up quark masses.  It can be seen that the latter 
interaction is of Majorana type, combining the 10 representation with 
itself.  

There is one global symmetry respected by these Yukawa interactions.  
If no Yukawa interactions exist, the theory is symmetric 
under phase rotations of each of the three fields $5^*_L$, $10_L$ and 
$5_H$.  The two Yukawa interactions break two of the three symmetries.  
The remaining U(1) has the following charge $X$:
\begin{equation}
X(5^*_L)=-3 \;,\quad
X(10_L)=1 \;,\quad
X(5_H)=-2 \:.
\end{equation} 
Although this ${\rm U(1)}_X$ is broken 
when the Standard Model Higgs field in $5_H$ develops a vacuum expectation 
value, the combination $X+4Y$ 
remains unbroken.  The combination $(X+4Y)/5$ turns out to be identical with 
$B-L$, the difference of baryon and lepton number.

The SU(5) GUT group can be elevated to a larger group SO(10) 
\cite{GeorgiInterview}, which has rank five, one rank larger than SU(5) 
and ${\rm SU(3)\otimes SU(2)\otimes U(1)}_Y$.  A single generation of 
quarks and leptons are unified into a single representation 16, 
which is a spinor representation of SO(10). 
This representation is a complex representation and is not equivalent to 
its conjugate $16^*$. 
 
The relation between SU(5) and SO(10) group can be understood from 
the decomposition of the fermion representations
\begin{equation}
16 = 5^* \oplus 10 \oplus 1 \;.
\end{equation}
There is an extra SU(5) singlet in 16.  This field is neutral under 
the Standard Model gauge group and can be naturally interpreted as 
the (charge conjugated) right-handed neutrinos.  The $X$ charge given above, 
with the assignment $X(1_L)=5$, becomes a gauge symmetry in SO(10), 
as the fifth diagonal generator: 
${\rm SO(10)} \supset {\rm SU(5)}\otimes {\rm U(1)}_X$.  
The fermion masses arise from the product 
\begin{equation}
\label{sixteensquared}
16 \otimes 16 = (10 \oplus 126)_S \oplus 120_A
\end{equation}
Higgs fields giving quark and lepton masses needs to be in a representation 
in the right-hand side of (\ref{sixteensquared}).  The Yukawa couplings 
are again in Majorana-type form.  
The SU(5) $5_H$ and its conjugate $5^*_H$ are in the vector 10 representation 
of SO(10): $10 = 5 \oplus 5^*$.

\section{Supersymmetry}

Principle of Relativity states that physics laws are the same in any inertial 
coordinate frame.  This means that the theory is invariant under the 
Poincar\'e group (inhomogeneous Lorentz group), whose generators are 
energy-momentum four vector $P^\mu$ and Lorentz generator $M^{\mu\nu}$ 
consisting of boost and angular momentum operators.  Conserved quantum 
numbers such as electric charge $Q$ are generators of internal symmetries 
which commute with the Poincar\'e group, and are Lorentz scalar.  
The question that if other kinds of symmetry can exist is negatively 
answered by the Coleman-Mandula theorem \cite{ColemanMandula}. 
With the assumption of nontrivial interactions (scattering matrix $S\ne1$) 
and massive particle states, vector charges or higher-rank tensor charges 
other than the Poincar\'e generators are not allowed.  

If one extends the concept of symmetry to include operations relating 
bosons to fermions, with anticommuting (Grassmann) parameters, new kinds 
of symmetry become possible.  The corresponding charges are restricted 
to be a spinor (having spin-1/2) \cite{HaagLS}, and the algebra 
of the spinor charges provide an extension of the Poincar\'e algebra.  
This is called supersymmetry \cite{WessBagger}.  
In its minimal version (with one kind of spinor charge, $N=1$), 
the spinor charge forms a Majorana spinor with its anticommutator given by
\begin{equation}
\{Q_\alpha, \bar Q_\beta\} = 2 (\gamma^\mu)_{\alpha\beta} P_\mu \;,
\end{equation}
or equivalently
\begin{equation}
\{Q_\alpha, Q_\beta\} = -2 (\gamma^\mu C)_{\alpha\beta} P_\mu \;.
\end{equation}
The Hamiltonian can be read off from these relations.
\begin{equation}
\begin{split}
H = P^0 
&= {1\over4}\bigl( \bar Q \gamma^0 Q + Q^T (\gamma^0)^T \bar Q^T \bigr)\\
&= {1\over4}\bigl( Q^\dagger Q + Q Q^\dagger \bigr) \;.
\end{split}
\end{equation}
The last expression explicitly shows that $H$ is a nonnegative operator.  
Supersymmetry is thus interwoven with Poincar\'e generators to form 
super Poincar\'e algebra.  It may be regarded as the only possible 
nontrivial extension of inhomogeneous Lorentz symmetry.  

The spectrum of the states in theories having these symmetries 
has special characteristics.  
To see the essence, it is useful to consider supersymmetry in $0{+}1$ 
dimensions, with the algebraic relation
\begin{equation}
H = Q^2
\end{equation}
Starting from a bosonic state $|B\rangle$, operating $Q$ to the state 
gives a fermionic state $|F\rangle=Q|B\rangle$.  If we apply $Q$ again, 
we find 
\begin{equation}
Q|F\rangle=Q^2|B\rangle=H|B\rangle=E_B|B\rangle
\end{equation}
The resulting state is nothing but the original bosonic state.  
Therefore the states form a pair made of a bosonic and a fermionic state.  
We can also see that the energy eigenvalues are the same for these two states 
by further operating $Q$ the third time.
When the first state is fermionic, the same conclusion is also obtained.  
Exception to this pairing can occur if $Q|B\rangle=0$.  Then the state 
$|B\rangle$ has zero energy (easily seen by applying $Q$ again).  
Thus physical states should always form a boson-fermion pair, except for 
zero energy states like the vacuum.

This situation also holds in higher space dimensions, and the number of 
bosonic and fermionic states have to be equal.  In our $3{+}1$ dimensions, 
The representation of the states depend on whether the particles are 
massless or massive.  For the massless case, states having 
helicities $\lambda$ and $\lambda+1/2$ form a pair.  The corresponding field 
multiplets containing the lowest spin states consist of a complex scalar and 
a Weyl spinor fields.  The degrees of freedom for bosons and fermions 
are both two.  This multiplet, called chiral supermultiplet, can also be 
written in terms of two real scalar and a Majorana field.  Next comes 
a pair of spin 1/2 and spin 1.  Helicities of the states are $\pm1/2$ 
and $\pm1$, the degrees of freedom are both two.  Existence of both 
positive and negative helicities is the consequence of CPT invariance.  
The massless spin-1 state is necessarily a gauge boson for consistent 
field theory description, and forms adjoint representation of the gauge 
group.  
The spin 1/2 fermion is a Majorana fermion and should also be in the adjoint 
representation.  

Massive multiplets are different from massless multiplets in general.  
This may be anticipated from the difference we have seen for Poincar\'e 
algebra.  Exception is the lowest spin case, spin 0 and spin 1/2. 
The smallest supermultiplet consists of two real scalars and a Majorana 
fermion.  If the fermion is charged, two complex scalars and a Dirac fermion 
make a massive supermultiplet, so the states must be doubled. 
A supermultiplet containing massive spin 1 states is more complicated.  
A massive spin 1 state has three helicity degrees of freedom and this 
should match with fermionic degrees of freedom.  One possibility is 
a supermultiplet containing one spin 0, two spin 1/2, and one spin 1 states 
(bosonic: $1+3$; fermionic: $2\times2$), which may be realized when the 
gauge symmetry is spontaneously broken by the vacuum expectation value of 
a scalar Higgs field.  
Another possibility is one spin 1/2, two spin 1, and one spin 3/2 states.

The equalness of bosonic and fermionic degrees of freedom has a 
profound consequence in field theories which possesses supersymmetry.  
A free field is equivalent to an infinite number of harmonic 
oscillators.  
Ground state of each oscillator has a zero-point energy of $\hbar\omega/2$ and 
the total vacuum energy diverges.  Usually the infinity is just subtracted 
in defining the energy of the vacuum.  
Situation is different in supersymmetric field theories.  
Each mode of fermions, unlike that of bosons, has a zero-point energy 
of $-\hbar\omega/2$ due to anticommuting property.  
When bosonic and fermionic degrees of freedom coincides with each other, 
as in supersymmetric theories, zero-point energies exactly 
cancel and the vacuum energy vanishes without subtraction.  
The absolute value of energy, therefore, is well-defined in 
supersymmetric theories.

\section{Minimal Supersymmetric Standard Model}

Cancellation between contributions from bosons and fermions in 
supersymmetric theory is not limited to the vacuum energy.  It has been 
invoked to resolve an uncomfortable situation 
related to the high energy structure of the Standard Model.  
One-loop correction to the Higgs self energy has quadratic divergence, 
which is absent in self energies of fermions and gauge bosons.  When we 
cut off the momentum integral over intermediate states around $\Lambda$, 
the one-loop self energy is of the order of $\lambda\Lambda^2/16\pi^2$, 
where $\lambda$ is the Higgs self coupling.  The cutoff $\Lambda$ should 
be larger than the limit under which scale the Standard Model is valid.  
If we expect the Standard Model to be valid up to the extremely large 
scale suggested by Grand Unification, the one-loop quadratic contribution 
becomes many order of magnitude larger than the physical Higgs mass term. 
Bare Higgs mass must be fine tuned extremely well (something like 
30 orders of magnitude) to cancel the loop contribution and leave 
the tiny physical $m_H^2$.  

This conceptual problem of fine tuning can be avoided if there is 
supersymmetry, because of cancellations of bosonic and fermionic 
contributions in the loop.  Cancellations occur not only in the 
vacuum energy (quartic divergence) but also in the quadratically 
divergent piece, leaving a logarithmically divergent piece which 
is not problematic.  
Of course, particles forming supersymmetric 
pairs do not exist in the Standard Model, and neither in the list of particles 
we have found so far.  To supersymmetrize the Standard Model requires 
introduction of `superpartners' for all particles in the Standard Model, 
doubling the number of degrees of freedom.  In addition, two Higgs doublets 
are necessary to produce masses of all quarks and leptons due to 
constraints on the interactions imposed by supersymmetry.  

The resulting theory is called the Minimal Supersymmetric Standard Model 
(MSSM) \cite{Drees, BaerTata}.  
It contains supersymmetric partners of every particle in the Standard Model, 
quarks and leptons, gauge bosons, and Higgs bosons.  These are named 
squarks and sleptons, gauginos, and higgsinos.  These particles have 
spin which differs by 1/2 from the original particle.  
The squarks and sleptons (sfermions) are scalars, 
and gauginos and higgsinos have spin 1/2.  
The internal quantum numbers of these superpartners are equal to 
those of the original particles.  

If the theory is totally supersymmetric, supersymmetry transformations 
commute with Hamiltonian, and supersymmetric pairs have equal masses.  
This is not the case in Nature, so supersymmetry must be broken.  
Spontaneous breaking of supersymmetry may be attractive, but do not work 
in MSSM, so explicit supersymmetry breaking terms are added to the Lagrangian. 
These breaking terms are restricted to the so-called `soft' breaking, 
which does not generate quadratic divergence and upset the cancellation 
in the Higgs self energy.  
This soft breaking is supposed to be the consequence of spontaneous 
breaking in a `hidden' sector contained in the whole theory, 
transmitted to our MSSM sector by some mechanism.
A number of such possibilities of `mediation' have been proposed.  
In addition, the masses of the superpartners should not be too large 
compared with the weak scale ($m_W$ or $m_H$), since the quadratic 
cutoff $\Lambda$ is replaced with the mass of superpartners in the loop 
integral.  

Among the Standard Model particles, gauge bosons (except $W^\pm$) are 
self-conjugate particles.   These gauge bosons, gluinos, photon, $Z$ 
have spin-1/2 superpartners which are self conjugate, Majorana fermions.  
Superpartners of the neutral Higgs bosons are also Majorana fermions.  
After the spontaneous breaking of the gauge ${\rm SU(2)\otimes U(1)}_Y$ 
symmetry due to the Higgs vacuum expectation values (the two Higgs 
fields both should develop vacuum expectation values), mixing arise between 
the neutral SU(2) and ${\rm U(1)}_Y$ gauginos and two higgsinos, 
giving four mass eigenstates in all.  
These mixed states are called neutralinos and all are Majorana particles.  
The lightest of these neutralinos is stable in most models, where 
discrete symmetry $R=(-1)^{3B+L+2J}$ is present.  All the known particles 
have $R=+$ and the superpartners have $R=-$, and the lightest neutralino 
is typically the lightest $R$-odd particle.  The lightest neutralino 
interacts with matter only weakly like neutrinos, and is expected to 
have a mass around the weak scale.  It is an 
attractive candidate for dark matter particles in the universe.  

Extensive searches have been made at high-energy colliders to discover 
various superpartners directly.  Unfortunately, no trace of supersymmetry 
has so far been found.  Proposals of possible pattern of supersymmetric 
particle masses compatible with present experimental limits are made.  
On the other hand, superstring theory, which unites all known fundamental 
interactions including gravity, needs supersymmetry for theoretical 
consistency, and we expect that supersymmetry appears 
at some energy scale, if not at the weak scale.

Motivations for supersymmetry have been rather theoretical.  
In contrast, we do have observational needs for physics beyond the 
Standard Model.  These include neutrino masses, dark matter, dark energy, 
baryon number of the universe, and inflation in very early universe.  
We have seen that at least three of them can be related to Majorana fermions.

\end{document}